\documentclass[12pt,preprint]{aastex}

\shorttitle{Disk Shadow Around NGC1333~ASR~41}
\shortauthors{Hodapp, Walker, Reipurth, Wood, Bally, Whitney, Connelley}

\received{2003 July 12}
\begin{document}

\title{A Disk Shadow Around the Young Star ASR~41 in NGC~1333}
\author{Klaus W. Hodapp\altaffilmark{1}, Christina H. Walker,\altaffilmark{2},
Bo Reipurth\altaffilmark{3}, Kenneth Wood\altaffilmark{2},
John Bally\altaffilmark{4}, Barbara A. Whitney\altaffilmark{5}, Michael Connelley\altaffilmark{1}}

\altaffiltext{1}{
Institute for Astronomy, University of Hawaii,\\
640 N. Aohoku Place, Hilo, HI 96720,
\\email: hodapp@ifa.hawaii.edu, connelley@ifa.hawaii.edu
}

\altaffiltext{2}{
School of Physics \& Astronomy, University of St. Andrews,\\
North Haugh, St. Andrews, Fife, KY16 9AD, Scotland,\\
email: cw26@st-andrews.ac.uk, kw25@st-andrews.ac.uk}

\altaffiltext{3}{
Institute for Astronomy, University of Hawaii,\\
2680 Woodlawn Drive, Honolulu, HI 96822,\\
email: reipurth@ifa.hawaii.edu
}

\altaffiltext{4}{
University of Colorado\\
Astrophysical \& Planetary Sci., 391 UCB, Boulder, CO 80309,\\
email: bally@origins.colorado.edu}

\altaffiltext{5}{
Space Science Institute\\
3100 Marine Street, Suite A353, Boulder, CO 80303,\\
email: bwhitney@colorado.edu}

\begin{abstract} 

We present images of the young stellar
object ASR~41 in the NGC~1333 star forming region 
at the wavelengths of H$\alpha$ and [SII] and in the
{\it I}, {\it J}, {\it H}, and {\it K}-bands. 
ASR~41 has the near-infrared morphology of an edge-on
disk object, but appears an order of magnitude larger than
typical systems of this kind. 
We also present detailed 
models of the scattering and radiative
transfer in systems consisting of a young star surrounded by
a proto-planetary disk, and the whole system being embedded
in either an infalling envelope or a uniform molecular
cloud. 
The best fit to the observed morphology can be achieved
with a disk of $\approx$200~AU diameter, immersed in a low
density cloud ($\approx$2$\times$10$^{-20}$ gcm$^{-3}$).
The low cloud density is necessary to stay below the sub-mm flux upper
limits and to preserve the shadow cast by the disk via single scattering.
The results demonstrate that ASR~41 is probably not
inherently different from typical edge-on disk
objects, and that its large apparent size is due to the
shadow of a much smaller disk being projected into the
surrounding dusty molecular material.

\end{abstract}


\section{Introduction}

Young low-mass stars form by accretion of material through a disk,
a process accompanied by outflow activity.
When most of the nascent
envelope has been dispersed in the later phases of star formation, 
a bipolar reflection nebula can be seen in the near infrared
if the object is observed in the plane of the disk. Under this condition the
star is obscured from direct view and therefore the total near-infrared flux
is dominated by the scattered light from material above and
below the disk plane.
Many examples of such edge-on disks have been found recently.
Most of the disks in those objects have radii inferred from
the optical or near-infrared morphology that are of the
order of magnitude of our solar system (i. e., $\approx$ 100 AU).
The prototypical HH~30~IRS
\citep{bur96} has a radius of $\approx$~250~AU, IRAS 04302+2247 \citep{luc98} has
a measured radius of $\approx$~200~AU in the near infrared, 
while the secondary component in the HK~Tauri
system \citep{sta98} has $\approx$~105~AU radius. The objects in the Taurus star
forming region \citep{pad99} range in radius from 300 to 500~AU and 
the object found by \citet{mon00} has about 60~AU radius.
In the Ophiuchus star forming region
\citet{bra00} found $\approx$150AU disk radii, 
while \citet{gro03} report a 300~AU radius and 
LkH$\alpha$263C \citep{jay02} has $\approx$ 150~AU radius. 

Increasingly sophisticated models for these objects have been
developed that treat multiple scattering, absorption and
radiative transfer by dust grains. These models 
reproduce many of the observed features
\citep{whi92,whi93,woo98,woo01,whi03,wol03}, 
constrain the density distribution around the
central star, and give information about the inclination of
the disk. 

In the course of a near-infrared study of the star
forming region NGC~1333 in the Per~OB2 complex, we found the
young star 
ASR~41 \citep{asp94} to show the morphology of an
edge-on disk object with two exceptionally large scattering
lobes. 
The edge-on disk structure of ASR~41 has been independently
discovered by \citet{els03} as part of their survey of giant
molecular clouds.
We assume NGC~1333 to be at the same distance of 316 pc
that \citet{her98} determined for the IC~348 cluster since they 
are both located in the same
Per~OB2 molecular cloud complex.

In this paper we present optical and infrared images of ASR~41
in combination with advanced scattering and 
radiative transfer models, and we argue that the unusual
size of this object (radius $\approx$ 3000 AU) 
is not indicative of the true extent of the
disk, but rather is a projection effect of the shadow
of a much smaller disk into the surrounding dusty molecular
cloud material.

\section{Observations and Data Reduction}

The NGC~1333 region was imaged with the QUIRC camera \citep{hod96}
at the UH~2.2m telescope, on the nights of Jan. 8 and 9, 2003 (UT),
with the intent of surveying this region for substellar mass objects. 
The night was photometric, and observations in the {\it J}, {\it H}, 
and {\it K} ``Mauna Kea Observatories'' (MKO) filter set
\citep{tok02} were obtained. The frames containing ASR~41 were part of
a larger mosaic. The area surrounding ASR~41 was covered by 30 individual
frames in each filter, the integration times were 150 s in {\it J}, 100 s in {\it H},
and 50 s in {\it K}, for total integration times of 4500 s in {\it J}, 3000 s in {\it H},
and 1500 s in {\it K}. Photometric calibration is based on 
the UKIRT standard FS111.
A difficulty is posed by the bright sources and extended emission in
this and all neighboring fields. The median filtering procedure to
establish the sky frames leaves small residuals at the location of the
bright sources, and contains large-scale gradients due to the presence
of extended emission. The faint outer wings of the extended flux 
surrounding ASR~41 are therefore subject to artifacts from poor sky
subtraction.

The H$\alpha$ (6569\AA/80\AA),  [SII] (6730\AA/80\AA),
and {\it I}-band (8220\AA/1930\AA) images were obtained Oct. 13 and 14, 2001,
using the 4 m Mayall telescope at NOAO with the 
MOSAIC CCD camera at a scale 0.26$"$/pix \citep{get02}.
{\it I}-band photometry of ASR~41 is based on the {\it I}-band magnitudes
of neighboring stars given by \citet{get02}.
Images taken earlier (Oct. 29, 1997) with the same equipment
were used to measure the proper motion of the newly discovered Herbig-Haro
object HH 727. 

\section{Results and Discussion}

Fig. 1 shows the {\it K}-band image of ASR~41 and its surroundings,
including the Herbig-Haro object HH~727, while
Fig. 2 shows ASR 41 (and HH~727 as an insert) in the H$\alpha$, [SII], {\it I}, 
{\it J}, {\it H}, and {\it K} bands. 
Going from the shortest wavelength covered here (H$\alpha$)
to the {\it J} band, the morphology of ASR~41 changes dramatically.
At H$\alpha$ ASR~41 appears as a faint patch of nebulosity
with a faint condensation at its center. At [SII], the central
source appears brighter and the extended emission begins to 
show the ``hourglass'' shape of a bipolar nebula.
The {\it I} band begins to show the bifurcation of the extended emission,
albeit less pronounced than at the longer near-infrared 
wavelengths. 
The dark band at an angle of about 137$^\circ$, bifurcating
the reflection nebulosity associated with ASR~41, 
is most prominent 
from the {\it J} to the {\it K} band.
The reflection nebula can be traced out to about 10$\arcsec$ 
from the center ($\approx$ 3000 AU) before
it gets confused with artifacts from the sky subtraction process.
The linear size of the 
scattering region therefore is about 10 to 20 times larger 
than in the typical edge-on disk systems listed in
the Introduction.
Either, ASR~41 is a unique object surrounded by a huge 
extended disk or the dark band is the shadow of 
a much smaller (and typical) disk which is projected into 
the dusty material surrounding the object, as is strongly
suggested by the wavelength dependence of its morphology.
If the dark band
were caused in its full extent by absorption in an edge-on dust disk in front of
the scattering region, one would expect the dark bifurcating band
to be more pronounced at shorter wavelengths. This is clearly not
the case. Rather, our images suggest that the dark band is the shadow of a
much smaller disk, projected into the surrounding dusty
medium. The transition from single scattering at longer
wavelengths, which preserves the shadow, to multiple
scattering at shorter wavelengths, which fills in the shadow, 
occurs at wavelengths around 1$\mu$m.

Within the dark band, a 
central object is clearly visible at all wavelengths from
[SII] to the {\it K} band. 
This central object has a FWHM of 0.69$\arcsec$ in the {\it K} band,
while the average FWHM of five other stars in Fig. 1 is
0.61$\pm$0.01$\arcsec$. The central object is thus marginally different
from a star; reflection nebulosity is contributing to its
FWHM.
Its position was measured in the {\it J}-band relative to 4 stars 
in the USNO-B catalog \citep{mon03} to be
3$^h$~28$^m$~51$^s$.3~+31$^\circ$~17$\arcmin$~40$\arcsec$~(J2000),
with estimated errors of $\approx$ 0.25$\arcsec$ for the absolute coordinates.
Astrometry of the central object in the [SII] image, using
a larger set of USNO-B catalog stars than in the case of the
infrared image, showed that it has the same coordinates
as the corresponding object in the {\it J}-band image
within the relative errors of $\approx$ $\pm$0.14$\arcsec$.
We don't see evidence for the wavelength dependent
position of the central brightness peak that is often found
in cometary or bipolar nebulae associated with Class I objects
that are dominated by the absorption by a large and massive
disk, e.g., \citep{hod88}.

The central source has magnitudes 
in an 1.5$\arcsec$$\times$1.5$\arcsec$
box centered on the central object of {\it I}= 20.2, {\it J}= 17.4, {\it H}= 16.1, {\it K}= 15.3.
The central source has a color of {\it H-K} = 0.73 and
{\it J-H} = 1.36 while the color of the reflection nebulosity
is bluer: 
We have taken measurements in 1.5$\arcsec$$\times$1.5$\arcsec$ boxes
at two additional positions to the north-east of the central star, i. e., in the
reflection nebula, as indicated
in Fig. 2.
In the additional box closer to the central object, we measured
{\it H-K} = 0.24 and {\it J-H} = 1.05, 
in the box farther to the north east we measure 
{\it H-K} = 0.34 and {\it J-H} = 1.15, with estimated
errors in these colors of $\approx$ $\pm$0.05 mags. 
This is consistent with the 
flux from the central star and possibly the innermost parts of the
reflection nebula being
reddened by absorption in the small disk, and the flux in 
the extended reflection nebula being bluer, possibly because
of preferential scattering by small dust particles.

ASR~41 is not associated with a strong sub-mm source. The 
object lies just at the southern tip of a ridge of extended 850$\mu$m
emission associated with the HH~12 complex \citep{san01}, but
it is not detectable as an individual point source in their map.
From the fact that even the lowest contour (75 mJy/beam)
of their map 
is not significantly distorted at the position
of ASR~41, we estimate an upper limit to its 850$\mu$m flux of
200 mJy.
This implies that
the total mass of dust involved in the scattering of light
from the central source, and the mass of the disk around the
central star, must be relatively small. This point will be discussed
in more detail in Section 4.

To the south-west of ASR~41, a faint extended object at
3$^h$~28$^m$~48$^s$.1~+31$^\circ$~17$\arcmin$~10$\arcsec$~(J2000) is
most prominent in H$\alpha$ and [SII] and
faintly visible in {\it K} (Fig. 2, inserts).
This is the photometric signature of a 
low excitation Herbig-Haro object without significant [FeII] emission
in the {\it H} band
\citep{rei01}. The object has therefore been named HH 727. 
Proper motion measurements on
the [SII] images from 1997 and 2001 show a small (about 1 pixel)
shift of the photocenter of the HH knot towards the south west
(PA = 215 $\pm$ 15). The shift corresponds to $\approx$ 100 km/s
in the general direction away from ASR~41, roughly perpendicular to
the plane of the bifurcating disk. While more precise
measurements are clearly desirable, this strongly suggests that
the HH object is physically associated with ASR~41.

\section{Modeling of the ASR~41 Disk Shadow}

We have computed detailed models for ASR~41
using a Monte Carlo scattered light code 
\citep{whi92}, \citep{whi93}; updated in \citep{whi03}. 
The models contain a number of simplifying assumptions,
in particular about the grain properties, and therefore
cannot be expected to reproduce all features of the observations.
The main purpose of the modeling is to demonstrate
that the object can be explained well 
within the current paradigm of star formation.
The models discussed here are based on the assumption 
that ASR~41 is essentially a Class II T-Tauri star surrounded 
by a disk of roughly the size of our solar system (r $\approx$ 100 AU), 
i.e., very similar to other edge-on disk systems. 

The two models differ in the assumptions about the
distribution of the scattering dust surrounding the central star
and its protoplanetary disk.
Two types of models were considered: disk plus constant 
density cloud (the ``disk + cloud model'') and disk plus infalling envelope
(the ``disk + envelope model''). 
Both reproduce the dark band as the shadow of a smaller disk.

The geometry usually used to model young bipolar nebulae
is an infalling envelope model, with the infall rate
adjusted to model objects in different evolutionary classes
\citep{whi93,whi03}.
For ASR~41, the infall rate is constraint
by the upper limit
for the sub-mm flux. 
The ``disk + envelope model'' in Fig. 3 uses an
outer disk radius of 200 AU,  a disk mass of 0.06 M$_{\odot}$,
an outer envelope radius of 7000 AU, and
an infall rate of 3.5$\times$10$^{-6}$M$_{\odot}$/yr 
that is similar to the rate used in other
successful models of young stellar objects \citep{whi93}.
The infalling envelope has the density distribution
of the free-fall models of \citet{ulr76} and \citet{cas81}. 
This model reproduces the basic features of the observations,
but shows a steeper intensity gradient with distance from
the illuminating object than was observed.
It predicts a 850$\mu$m flux of 71 mJy,
consistent with the sub-mm upper limit.

The other model is based on the assumption outlined in
qualitative form earlier that ASR~41 consists of a typical
small edge-on disk object and a surrounding scattering cloud
of uniform density
into which the disk shadow gets projected. For this model,
the density of the surrounding scattering material was
adjusted so that single scattering dominates, which preserves
the shadow effect and also coincidentally keeps the sub-mm flux within the
observed upper limit.
The ``cloud model'' uses a disk radius of 100 AU and a mass
of 0.005 M$_{\odot}$, for the best fit to the {\it K}-band data,
the cloud itself is assumed to have a radius of 10000 AU.
Our ``disk + cloud model'' assumes a cloud density of 2$\times$10$^{-20}$ gcm$^{-3}$,
that is within the range of typical densities in large molecular clouds.
This model produces a 850$\mu$m flux of 40mJy, well below the
sub-mm flux limit.
Increasing both the cloud density and disk mass by an order of
magnitude leads to slightly better fits to the near-infrared
fluxes, but predicts a 850$\mu$m flux of 234mJy, i. e., above
the flux limit.

Fig. 3 shows the {\it K}-band image of ASR~41 
(top, rotated to have the polar axis vertical for comparison with 
the models), and the two model flux distributions. 
Some noise was added to the model images to make them more 
visually comparable to the real image. 
The "disk + envelope" model is successful at matching
the photometry, but morphologically,
the flux is falling off too 
rapidly with distance from the central source due 
to the density distribution in the envelope. 
The ``disk + cloud model''
matches the near-infrared
morphology better, including the large extent of the
reflection nebulosity.

Both model fits to the observed photometry indicate inclinations of about 80$^\circ$,
even though the match to the observed morphology, in
particular the difference in flux from the two lobes
could probably be improved by assuming somewhat
lower inclinations.
These problem our models have with precisely 
matching the near-infrared photometry and morphology of the object point to 
shortcomings in the dust model used here. 
Our code is currently limited to one dust model throughout, 
and given the extent of the object an interstellar 
grain model was chosen. 
This is a good approximation for the particles expected 
in a thin molecular cloud, but may not be a good model 
for the grains in a much denser, proto-planetary 
dust disk in the close vicinity of a young star,
e.g. \citep{bec90,woo02}.
Due to the obvious simplifications in our model, they
are not suitable to derive the luminosity or detailed
information about the evolutionary state of the star
at the center of ASR~41.

\section{Summary}

We have presented images of the young stellar object ASR~41 in NGC~1333
at wavelengths ranging from H$\alpha$ to the {\it K}-band. The bipolar
nebula seen at the longer of these wavelengths is much larger than
typical edge-on disk systems but can be understood as the
shadow of a smaller disk being projected into the dusty
material of the surrounding molecular cloud. Detailed model
calculations of the scattering and radiative transfer in this
object have been presented. These models show that the morphological
features and the near-infrared fluxes
can be modeled by a small, low-mass disk and surrounding scattering
medium without violating the upper limit on the sub-mm flux.
ASR 41 is, most likely, a Class II T Tauri star with a disk
of roughly the size of our solar system, and overall similar
to other edge-on-disk objects. It is distinguished from those
objects by the scattering of the disk shadow in the surrounding
dusty molecular material.

\acknowledgements
NOAO is operated by the Association of Universities for Research in Astronomy (AURA),
Inc. under cooperative agreement with the National Science Foundation.

\clearpage
\begin{figure}
\figurenum{1}
\epsscale{0.8}
\plotone{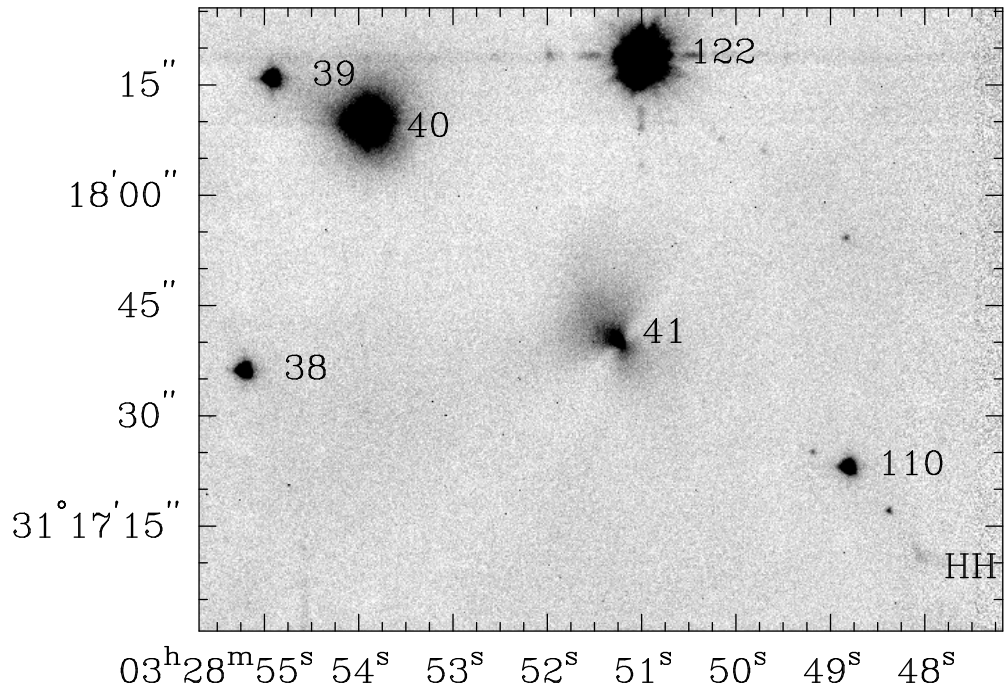}
\caption{
{\it K}-band image of ASR~41. The objects are labeled by ASR number \citep{asp94}.
The position of the Herbig-Haro object 727 is labeled HH.
}
\end{figure}

\clearpage
\begin{figure}
\figurenum{2}
\epsscale{0.8}
\plotone{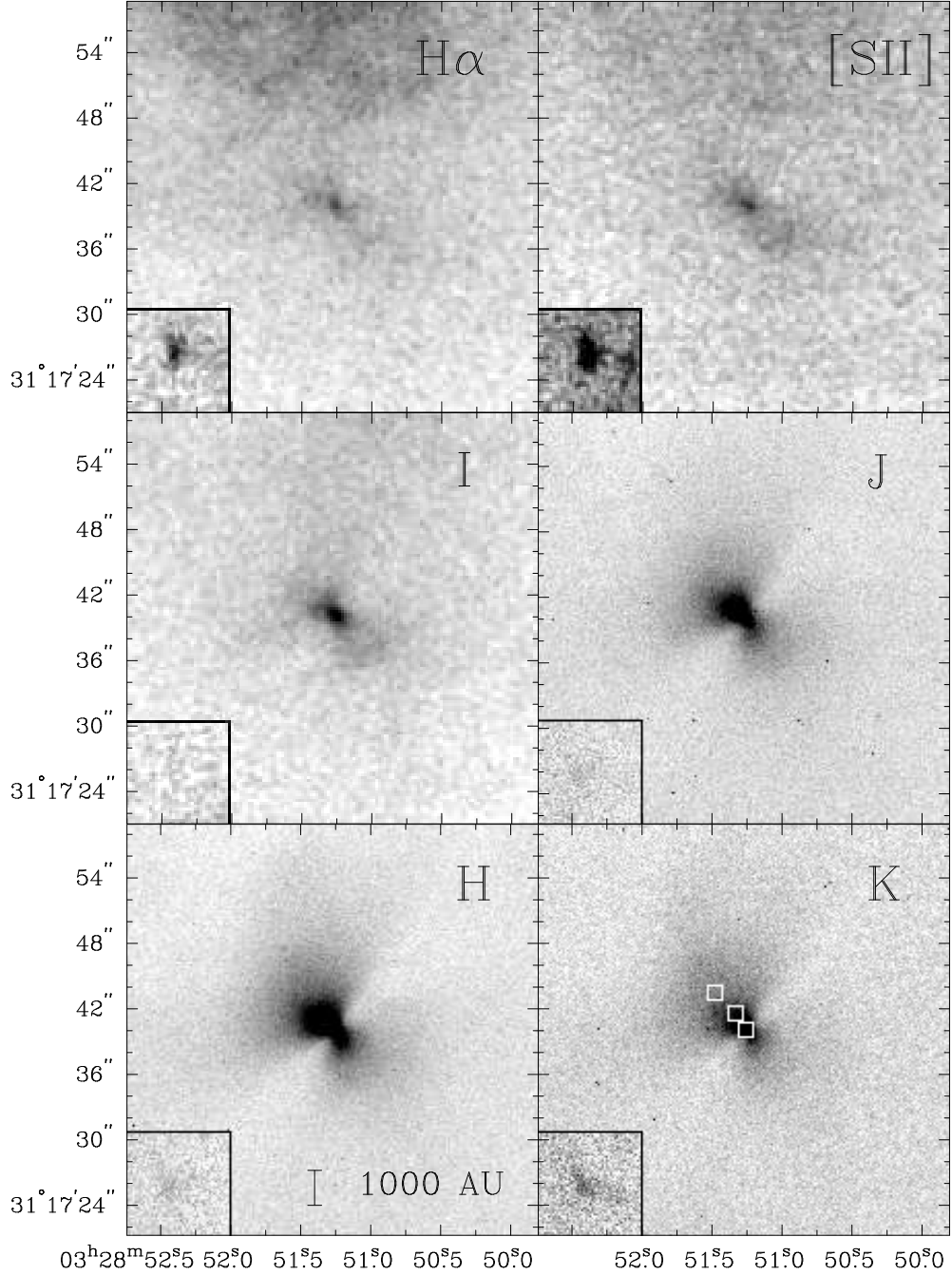}
\caption{
H$\alpha$, [SII], {\it I}, {\it J}, {\it H}, and {\it K}-band images of ASR~41.
Inserted in each image is a smaller image of the HH-object 727, at the same spatial
scale, but with the intensity scale stretched by a factor of two.
The observed morphology of ASR~41 changes dramatically from H$\alpha$ and [SII], where an
extended nebula with a central, almost pointlike source is visible,
to the {\it I} band, where the bifurcating dark band becomes noticeable. The bipolar
nature of the object and the full extent of the dark band are fully established in the 
{\it J} band, and the morphology then remains virtually identical in the {\it H} and
{\it K} bands. The inserted image of the HH object 727 is strongest in H$\alpha$ and [SII],
and indicated in K (due to H$_2$ line emission), but is not detectable or very weak in
the {\it I}, {\it J}, and {\it H} bands. 
Details of the morphology and photometry are discussed in Section 3.
}
\end{figure}

\clearpage
\begin{figure}
\figurenum{3}
\epsscale{0.8}
\plotone{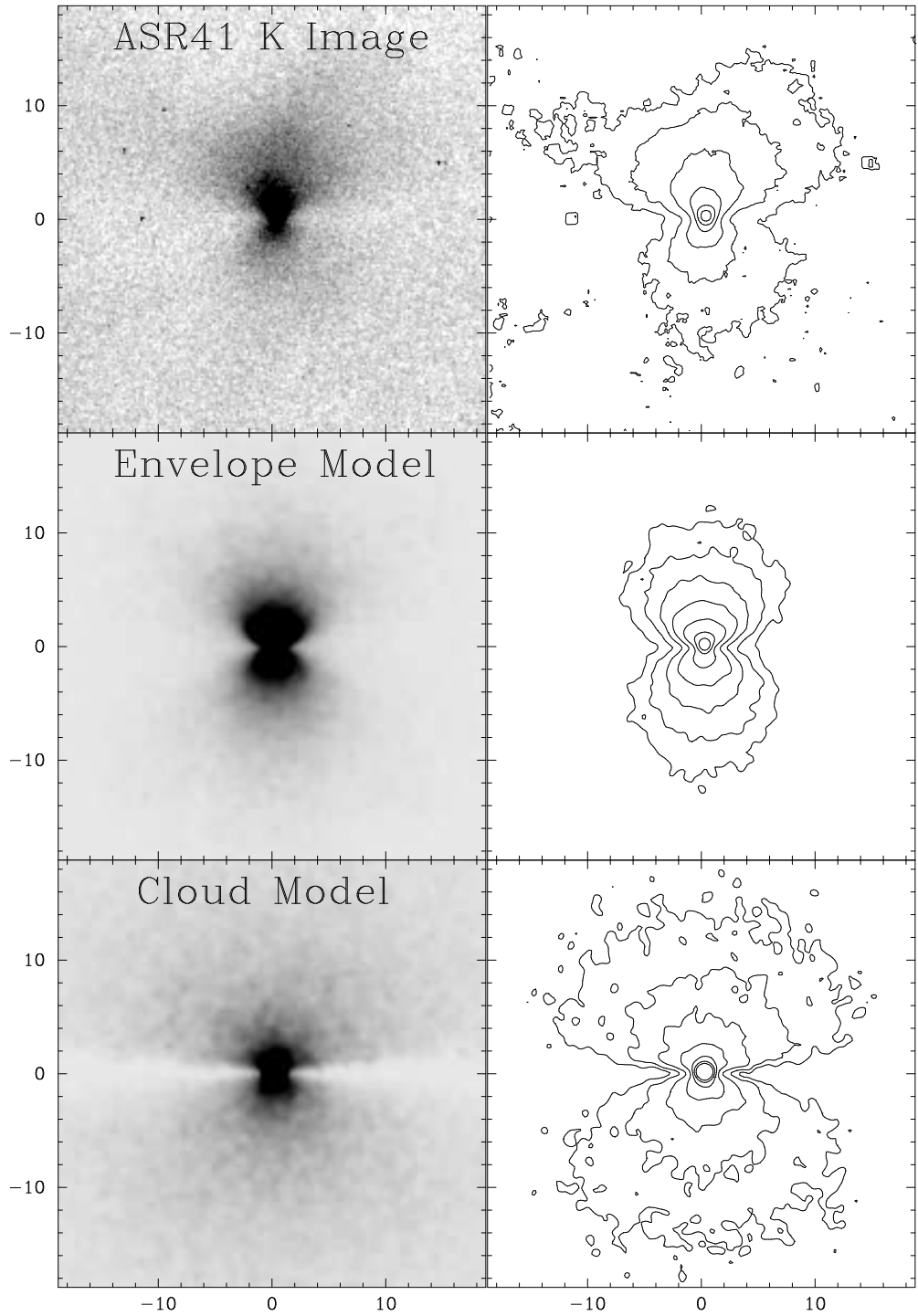}
\caption{
Comparison of the {\it K}-band image (rotated) of ASR~41 (top) with a ``disk + envelope model''
(center), and a ``disk + cloud'' model (bottom).
}
\end{figure}


\end{document}